\documentclass[
11pt,%
tightenlines,%
nofootinbib,preprintnumbers,prd]{revtex4}
\newcommand{\be}{\begin{equation}}
\newcommand{\ee}{\end{equation}}
\newcommand{\bea}{\begin{eqnarray}}
\newcommand{\eea}{\end{eqnarray}}

\newcommand{\0}{\over }

\begin{document}
\preprint{TUW-03-15}

\title{Gauge dependence identities for color superconducting QCD}
\author{A. Gerhold}
\affiliation{Institut f\"ur Theoretische Physik, Technische
Universit\"at Wien, Wiedner Haupstr.~8-10,
A-1040 Vienna, Austria }
\author{A. Rebhan}
\affiliation{Institut f\"ur Theoretische Physik, Technische
Universit\"at Wien, Wiedner Haupstr.~8-10,
A-1040 Vienna, Austria }

\begin{abstract}
Using generalized Nielsen identities a formal proof is 
given that the fermionic quasiparticle dispersion 
relations in a color superconductor are
gauge independent. This turns out to involve gluonic tadpoles
which are calculated to one-loop order in a two-flavor
color superconductor. Regarding the appearance of gluon tadpoles,
we argue that in QCD
the color superconducting phase is automatically color neutral. 
\end{abstract}
\maketitle

\section{INTRODUCTION}
It is well known that sufficiently cold and dense quark matter is a color superconductor 
\cite{Bailin:1984bm,Rajagopal:2000wf
}.
At asymptotic densities, when the QCD coupling constant is small, it is
possible to investigate the properties of a color superconductor using mostly perturbative
techniques based on the fundamental Lagrangian of QCD, because the
dominant interaction leading to the formation of Cooper pairs
is single-gluon exchange, which is attractive in the color-antitriplet
channel. The correct leading order result for the gap
has been first obtained by Son \cite{Son:1998uk},
\begin{equation}
  \phi={b_0\over g^5}\mu\,\exp\left(-{c\over g}\right)[1+\mathcal{O}(g)],
\end{equation}
with $c=3\pi^2/\sqrt 2$. 
This result has been confirmed, and the constant $b_0$ has been determined \cite{Pisarski:1999tv,
Schafer:1999jg,Hsu:1999mp,Hong:1999fh}, 
taking into account also contributions from the quark self energy \cite{Brown:1999aq,Wang:2001aq}.

The gap equation is usually derived from the Schwinger-Dyson equation, where some approximations
are made such as the neglect of vertex corrections. 
In \cite{Rajagopal:2000rs} it has been shown that large gauge dependences occur in such calculations
of the value of the gap for $g>g_c\sim0.8$. 
In \cite{Pisarski:2001af} it has been shown that the corresponding
gauge dependent terms appear in the
gap equation at sub-sub-leading order, though gauge dependences arise
already at sub-leading order if the gap parameter is not evaluated on
the quasiparticle mass shell.
Recently, the gap equation has been 
considered 
in a non-local gauge where the quark self energy and vertex corrections vanish \cite{Hong:2003ts} 
to include higher-order corrections to the prefactor,
exploiting the expected gauge independence of the gap as a physical quantity.


In view of all these efforts, 
it seems desirable to have a formal proof 
of gauge independence of the relevant quantities. In this Note we
shall consider
the fermionic quasiparticle dispersion relations in a color superconductor,
which are modified by the appearance of a gap which dynamically breaks
color symmetry. 
Such a proof is given in
section 2 of the present paper as a straightforward generalization of a similar proof
for finite temperature QCD \cite{KKR}. It turns out that in order to formulate this proof
one 
has to allow for the appearance of gluon tadpoles 
in a color superconducting phase. We demonstrate
the possible existence of such tadpoles in section 3 by
a one-loop calculation in the example of an $N_f=2$
color superconductor, discussing also briefly the issue of
color neutrality. We argue that in QCD (in contrast
to NJL models) the color superconducting
phase is automatically color neutral.

Our notations follow mostly those of
Pisarski and Rischke, Refs.~\cite{Pisarski:1999tv,Pisarski:1999av,Rischke:2000qz}.

\section{GAUGE INDEPENDENCE OF FERMIONIC DISPERSION RELATIONS}

We consider the inverse quark propagator in the Nambu-Gor'kov formalism
\cite{Bailin:1984bm,Wang:2001aq,Manuel:2000nh},
\begin{equation}
  \mathcal S^{-1}=\left(
    \begin{array}{cc} Q\!\!\!\!/+\mu\gamma_0
    +\Sigma & \Phi^- \\
    \Phi^+ & Q\!\!\!\!/-\mu\gamma_0+\bar\Sigma 
    \end{array}
  \right), \label{e1}
\end{equation}
where $\Phi^\pm$ are the gap functions, related by
$\Phi^-(Q)=\gamma_0 [\Phi^+(Q)]^\dag\gamma_0$, and 
$\bar\Sigma(Q)=C[\Sigma(-Q)]^T C^{-1}$ with the charge 
conjugation matrix $C$.
Flavor and fundamental color indices are suppressed in (\ref{e1}). This inverse propagator
is the momentum space version of the second derivative of
the effective action,
\begin{equation}
  {\delta^2\Gamma\over\delta\bar\Psi(x)\delta\Psi(y)}\Big|_{\psi=\bar\psi=A_i^a=0,
  A_0^a=\tilde A_0^a},
  \label{e2}
\end{equation}
where $\Psi=(\psi, \psi_c)^T$, $\bar\Psi=(\bar\psi, \bar\psi_c)$, and $\tilde A_0^a$
is the expectation value of $A_0^a$, which we allow to be non-vanishing (see section 3).

It should be noted that the doubling of fermionic fields in
terms of $\Psi$ and $\bar\Psi$ is just a notational convenience here;
the effective action itself should be viewed as depending only
on either $(\psi,\bar \psi)$ or the set $\Psi=(\psi, \psi_c)^T$.

The gauge dependence identity for the effective action (generalized Nielsen identity, see appendix) follows from considering a completely arbitrary
variation of the gauge fixing function. It
can be written as
\begin{eqnarray}
  &&\!\!\!\!\!\!\!\!\!\!\!\!\!\!\!\!\delta\Gamma =\int dx\left({\delta\Gamma\over\delta\psi(x)}
  \delta X_{(\psi)}(x)-\delta X_{(\bar\psi)}(x)
  {\delta\Gamma\over\delta\bar\psi(x)}+{\delta\Gamma\over\delta A^{a\mu}(x)}
  \delta X_{(A)}^{a\mu}(x)\right) \nonumber\\
  &&\!\!\!\!\!\!\!\equiv\int dx\left({\delta\Gamma\over\delta\psi(x)}
  \delta X_{(\psi)}(x)+{\delta\Gamma\over\delta\psi_c(x)}
  \delta X_{(\psi_c)}(x)+{\delta\Gamma\over\delta A^{a\mu}(x)}
  \delta X_{(A)}^{a\mu}(x)\right),\quad\label{e3}
\end{eqnarray}
where the various $\delta X$ are defined in (\ref{dGdX}).

Eq. (\ref{e3}) can be cast in a more compact form using the DeWitt
notation for the fermions,
\begin{equation}
  \delta\Gamma= \Gamma_{,i}\delta X^i+\int dx{\delta\Gamma\over\delta A^{a\mu}(x)}
  \delta X_{(A)}^{a\mu}(x), 
  \label{e5}
\end{equation}
where $i=(\psi(x), \psi_c(x))^T$, $\bar i=(\bar\psi(x), \bar\psi_c(x))$,
and the comma denotes functional derivation. Taking the second derivative of
(\ref{e5}), setting $\psi=\bar\psi=A^a_i=0$, $A^a_0=\tilde A^a_0$,
and using the fact that
\begin{equation} \label{gdiip1}
  {\delta\Gamma\over\delta A^a_0}\Big|_{\psi=\bar\psi=A^a_i=0,A^a_0=\tilde A^a_0}=0,
\end{equation}
we obtain a gauge dependence identity for the inverse propagator (\ref{e2}),
\begin{equation}
  \delta\Gamma_{i\bar j}=-\Gamma_{k\bar j}\delta X^k_{\,,i}+
  \Gamma_{ki}\delta X^{k}_{\,,\bar j}+\int dx{\delta\Gamma_{i\bar j}\over\delta A^{a0}(x)}
  \delta X_{(A)}^{a0}(x).
\end{equation}

Up to this point, our functional relations are completely general
and apply also to the case of inhomogeneous condensates which
may be realized in the so-called LOFF phase (see \cite{Casalbuoni:2003wh}
for a recent review). In this Note we do not attempt to cover
the complications this case may add to the question of gauge
independence, but continue by assuming translational invariance.
This allows us to introduce $\delta\tilde A^{a0}:=-\delta X_{(A)}^{a0}(x=0)$
and to write (\ref{gdiip1}) as
\begin{equation}
  \delta\Gamma_{i\bar j}=-\Gamma_{k\bar j}\delta X^k_{\,,i}-
  \delta X^{\bar k}_{\,,\bar j}\Gamma_{i\bar k}-{\partial\Gamma_{i\bar j}\over\partial \tilde A^{a0}}
  \delta\tilde A^{a0}.
  \label{e6}
\end{equation}
Furthermore, we can transform eq. (\ref{e6}) into momentum space,
\begin{equation} 
  \delta\Gamma_{i\bar j}(Q)+\delta\tilde A^{a0}{\partial\Gamma_{i\bar j}(Q)\over\partial \tilde A^{a0}}
  =-\Gamma_{k\bar j}(Q)\delta X^k_{\,,i}(Q)-
  \delta X^{\bar k}_{\,,\bar j}(Q)\Gamma_{i\bar k}(Q), \label{e7}
\end{equation}
where the indices $i$ and $\bar i$ from now on comprise only color, flavor, 
Dirac and Nambu-Gor'kov indices. Using the fact that 
\begin{equation}
  \delta\det M=(\det M)\,{\rm Tr}[M^{-1}\delta M]
\end{equation}
for any matrix $M$, we obtain from eq. (\ref{e7})
\begin{equation}
  \delta\det(\Gamma_{i\bar j})+\delta\tilde A^{a0}
  {\partial\over\partial \tilde A^{a0}}\det(\Gamma_{i\bar j})
\equiv \delta_{\rm tot}\det(\Gamma_{i\bar j})
  =-\det(\Gamma_{i\bar j})[\delta X^k_{\,,k}
  +\delta X^{\bar k}_{\,,\bar k}]. \label{e8}
\end{equation}
The left hand side of this identity is the total variation \cite{Nielsen:1975fs,Aitchison:1984ns} of the determinant of the inverse
quark propagator, with the first term
corresponding to the explicit variation of the gauge fixing function, and the second term coming from the
gauge dependence of $\tilde A_0^a$.

Since the determinant is equal to the product of the eigenvalues, eq. (\ref{e8}) implies 
that the location of the singularities of the quark propagator 
is gauge independent, provided the singularities of $\delta X^k_{\,,k}$ do not coincide
with those of the quark propagator. As in \cite{KKR} one may argue that $\delta X$ is
1PI up to a full ghost propagator, and up to gluon tadpole insertions
(see Ref.~\cite{KKR} for the explicit diagrammatic structure).
The singularities of the ghost propagator are not
correlated to the singularities of the quark propagator. 
Their fundamental difference is in fact enhanced by medium effects---e.g.,
there exist no hard-thermal/dense-loop (HTL/HDL) \cite{LeB:TFT}
corrections to vertex functions involving ghosts,
whereas all physical degrees of freedom acquire HTL/HDL self-energy
corrections.

Gauge independence of the zeros of the inverse fermion propagator
then follows provided that
also the 1PI parts of $\delta X$ have
no singularities coinciding with the singularities of the propagator.
An important caveat in fact comes from massless poles in the
unphysical degrees of freedom of the gauge boson propagator,
which are typical in covariant gauges and which can give rise
to spurious mass shell singularities as encountered in the case
of hot QCD in \cite{Baier:1992dy}.
But, as was pointed out in \cite{Rebhan:1992ak},
these apparent gauge dependences are avoided if the quasiparticle
mass-shell is approached with a general infrared cut-off such
as finite volume, and this cut-off lifted only in the end, i.e.,
after the mass-shell limit has been taken.
The tadpoles do not introduce any singularities.

The determinant which appears in (\ref{e8}) 
is taken with respect to color, flavor,
Dirac and Nambu-Gor'kov indices. The determinant in Nambu-Gor'kov space may be evaluated
explicitly by noting that
\begin{equation}
  \det\left(
    \begin{array}{cc}\mathcal{A}&\mathcal{B}\\ \mathcal{C}&\mathcal{D} 
    \end{array}
  \right)=\det(\mathcal{DA}-\mathcal{DBD}^{-1}\mathcal {C})
\end{equation}
for arbitrary $n\times n$ matrices $\mathcal A$, $\mathcal B$, $\mathcal C$, 
$\mathcal D$ (with $\mathcal D$ invertible). 
Hence we obtain from (\ref{e1})
\begin{eqnarray}
  \det(\Gamma_{i\bar j})&=&\det\big[(Q\!\!\!\!/-\mu\gamma_0+\bar\Sigma)
  (Q\!\!\!\!/+\mu\gamma_0+\Sigma)\nonumber\\
  &&\qquad-(Q\!\!\!\!/-\mu\gamma_0+\bar\Sigma)
  \Phi^-(Q\!\!\!\!/-\mu\gamma_0+\bar\Sigma)^{-1}\Phi^+
  \big].\qquad \label{e10}
\end{eqnarray}
The inverse of the matrix of which the determinant is taken here appears, of course, in the ordinary
quark propagator, which is obtained by inverting (\ref{e1}),
\begin{eqnarray}
  &&\!\!\!\!\!\!\!\!\!G^+(Q)=\big[(Q\!\!\!\!/-\mu\gamma_0+\bar\Sigma)
  (Q\!\!\!\!/+\mu\gamma_0+\Sigma)\nonumber\\
  &&\qquad\qquad-(Q\!\!\!\!/-\mu\gamma_0+\bar\Sigma)
  \Phi^-(Q\!\!\!\!/-\mu\gamma_0+\bar\Sigma)^{-1}\Phi^+
  \big]^{-1}(Q\!\!\!\!/-\mu\gamma_0+\bar\Sigma).\qquad
\end{eqnarray}

At leading order, when the quark self energy can be neglected, 
eqs. (\ref{e8}) and (\ref{e10}) imply that the gap function is gauge independent
on the quasiparticle mass shell.
It should be noted, however, that at higher orders only dispersion
relations obtained from (\ref{e10}), which also include the quark self energy $\Sigma$,
can be expected to be gauge independent.


\section{GLUON TADPOLE AND COLOR NEUTRALITY}

As we have seen, the gauge dependence identities (\ref{e8}) involve
gluon tadpoles in a non-trivial manner.

In a color superconductor global color symmetry is dynamically broken by the
diquark condensate. This means that there is no symmetry
which forbids the existence of gluon tadpoles.
Let us consider the one-loop tadpole diagram,
\begin{equation}
  \mathcal{T}^a=-{g\over2}\int {d^4Q\over i(2\pi)^4}\mathrm{Tr}_{D,c,f,NG}[\hat\Gamma_0^a
  \mathcal S(Q)]. \label{a1}
\end{equation} 
Here $\mathcal S(Q)$ is the quark propagator in the Nambu-Gor'kov basis, whose inverse
is given by (suppressing color and flavor indices) 
\begin{equation}
  \mathcal S^{-1}=\left(
    \begin{array}{cc} Q\!\!\!\!/+\mu\gamma_0 & \Phi^- \\
    \Phi^+ & Q\!\!\!\!/-\mu\gamma_0\end{array}
  \right) \label{a2}
\end{equation}
with $\Phi^-=\gamma_0(\Phi^+)^\dag\gamma_0$.
The quark-gluon vertex $\hat\Gamma_0^a$ is given by 
\begin{equation}
  \hat\Gamma_0^a=\left(\begin{array}{cc}\Gamma_0^a &0\\0&\bar\Gamma_0^a
  \end{array}\right),
\end{equation}
with $\Gamma_0^a=\gamma_0 T^a$ and $\bar\Gamma_0^a=-\gamma_0 (T^a)^T$.
The trace in (\ref{a1}) has to be taken with respect to Dirac, color, flavor and 
Nambu-Gor'kov indices.

The gap $\Phi^+$ is a matrix in Dirac, color and flavor space. We 
consider as an example an $N_f=2$ color superconductor and
take the 
following ansatz 
for the gap \cite{Pisarski:1999tv,Alford:1998zt}: 
\begin{equation}
  \Phi_{fg,ij}^+(Q)=\varepsilon_{fg}\varepsilon_{ij3}(\phi^+(\mathcal{P}^+_{r+}-\mathcal{P}^+_{l-})
 +\phi^-(\mathcal{P}^-_{r-}-\mathcal{P}^-_{l+})),
\end{equation}
where $f$, $g$ are flavor indices and $i$, $j$ are fundamental color indices. 
The $\mathcal P$'s are the projection operators introduced in \cite{Pisarski:1999av}.
We have  assumed for simplicity that the right-handed and left-handed gap functions are 
equal up to a sign \cite{Pisarski:1999tv}. 

First we evaluate the trace over Nambu-Gor'kov space which gives
\begin{equation}  
  \mathcal{T}^a=-{g\over2}\int {d^4Q\over i(2\pi)^4}\mathrm{Tr}_{D,c,f}[\Gamma_0^a
  \mathcal{G}^+(Q)+\bar\Gamma_0^a\mathcal{G}^-(Q)],
\end{equation}
where $\mathcal{G}^\pm$ is obtained from inverting (\ref{a2})\cite{Rischke:2000qz},
\begin{equation}
  \mathcal{G}^\pm_{fg,ij}=\delta_{fg}[(\delta_{ij}-\delta_{i3}\delta_{j3})G^\pm
  +\delta_{i3}\delta_{j3}G_0^\pm],
\end{equation}
with \cite{Rischke:2000qz}
\begin{eqnarray}
  G^\pm(Q)&=&\sum_{e=\pm}{q_0\mp(\mu-eq)\over q_0^2-(\mu-eq)^2-|\phi^e|^2}\Lambda^{\pm e}
  ({\bf q})\gamma_0,\\
  G^\pm_0(Q)&=&\sum_{e=\pm}{q_0\mp(\mu-eq)\over q_0^2-(\mu-eq)^2}\Lambda^{\pm e}
  ({\bf q})\gamma_0,
\end{eqnarray}
where $\Lambda^\pm$ are energy projectors, 
$\Lambda^\pm({\bf q})={1\over2}(1\pm\gamma_0\gamma^i\hat q^i)$. 
Evaluating the trace over flavor and color space 
we get
\begin{equation}  
  \mathcal{T}^a=g(T^a)_{33}\int {d^4Q\over i(2\pi)^4}\mathrm{Tr}_D
  [\gamma_0 (G^+-G^-- G_0^++G_0^-)].
\end{equation}
Assuming that $\phi^-\simeq0$ \cite{Pisarski:1999tv} and that $\phi^+$
has negligible four-momentum dependence in the vicinity of
the quasiparticle pole we obtain
\begin{eqnarray}
  \mathcal{T}^a&\simeq&-4g(T^a)_{33}\int {d^4Q\over i(2\pi)^4}\left({\mu-q\over q_0^2-(\mu-q)^2
  -|\phi^+|^2}-{\mu-q\over q_0^2-(\mu-q)^2}
  \right)\nonumber\\
  &\simeq&{g\over\pi^2}(T^a)_{33}\int_0^\infty dq\,q^2
  (\mu-q)\left({1\over\sqrt{(\mu-q)^2+|\phi^+|^2}}-{1\over|\mu-q|}\right).
\end{eqnarray}
In order to obtain an order of magnitude estimate, we make the approximation 
$\phi^+(q)\simeq\phi^+_0\theta(2\mu-q)$ 
with $\phi^+_0=const.$ \cite{Rischke:2000qz}. 
Then the $q$-integration can be readily performed, with the result
\begin{equation}
  \mathcal{T}^a\simeq-{2g\over\pi^2}(T^a)_{33}\, \mu\,(\phi_0^+)^2
  \,\mathrm{ln}\left({\phi^+_0\over2\mu}\right)
  +{\mathcal O}((g\mu\phi_0^+)^2),
\quad (T^a)_{33} = -\delta^{a8}/\sqrt3.
 \label{e27}
\end{equation}
This result is in fact of order $\mu\phi^2$, 
because $\mathrm{ln}(\phi/(2\mu))$ is of order $1/g$.
At this order it cannot be excluded that there are cancellations from higher-loop contributions which have 
been neglected
so far \cite{Rischke:2000qz}. In particular it is still possible that $\mathcal{T}^a\equiv0$, although this seems
to be unlikely as there is no symmetry which forces $\mathcal{T}^a$ to be exactly zero.

To address 
the question of color neutrality \cite{Amore:2001uf,Alford:2002kj,Steiner:2002gx,Shovkovy:2003uu} we consider 
the partition function
\begin{equation}
  \exp(-\Omega/T)=\int\mathcal{D}\varphi\,\exp(-S[\varphi]),
\end{equation}
where $\varphi$ denotes the set of all fields, and $S[\varphi]$ is the QCD action (including gauge 
fixing terms and ghosts). Following the argument given 
in \cite{Khlebnikov:1996vj} it is easy to see that the system described by this 
partition function is color neutral, at least if one chooses a gauge fixing which
does not involve $A_0^a$, for instance Coulomb gauge:

The fields $A_0^a$ appear in the action as Lagrange multipliers for the Gauss law constraint
\cite{ItzZ:QFT}.
 Therefore in the path integral the integration over the zero-momentum modes $A_{0,\vec p=0}^a$  
produces delta functions, $\delta(N_a)$, where $N_a$ are the color charges. This means that
only color neutral field configurations contribute to the partition function, q.e.d.

The gluon tadpole obtained in (\ref{e27}) 
may in fact be interpreted as an {\em effective} 
chemical potential for the
color number eight of the order 
$$ g m_D^{-2} \mu \phi^2 \sim \phi^2/(g\mu), $$
where $m_D\propto g\mu$ is the corresponding leading-order Debye mass
\cite{Rischke:2000qz}.
It may be noted that the chemical potential $\mu_8$ which has been found by requiring
color neutrality in an NJL model is also proportional to $\phi^2$ \cite{Steiner:2002gx}. We emphasize however that whereas
in NJL models color neutrality has to be imposed as an additional condition,
color neutrality is guaranteed automatically in QCD by the integration over the $A_0^a$ 
zero-momentum modes.

\section{CONCLUSIONS}

We have presented a formal proof of gauge independence for
the fermionic quasiparticle dispersion relations in color superconducting QCD.
As long as the quark self energy can be neglected, this implies
gauge independence of the gap function on the quasiparticle mass shell.
In general, however, only the singularities of the
propagator, which involve also the gauge-dependent quark self energy,
can be expected to be gauge independent. At higher orders,
gauge parameter variations may also affect the temporal
component of the gluon tadpole, which, as we have shown in
a simple example, can be nonvanishing in a color superconductor.
While a nonvanishing gluon tadpole may be viewed as an effective
chemical potential for color charge, we have argued that
color neutrality is an automatic consequence of QCD, if no
explicit chemical potentials for color are introduced.

\acknowledgments

We would like to thank R.~Pisarski
and D.~Rischke for useful correspondence.
This work has been supported by the 
Austrian Science Foundation FWF, project no.~P16387-N08.

\appendix
\section{Gauge dependence identity for the effective action}

An arbitrary gauge theory is defined by an action $S_{inv}[\varphi]$ which is invariant under 
(infinitesimal) gauge transformations\footnote{We use the DeWitt notation, where a Latin index
comprises all discrete and continuous field labels, and a Greek index comprises group and space-time
indices.},
\begin{equation}
  \delta\varphi^i=D^i_{\,\alpha}[\varphi]\delta\xi^\alpha.
\end{equation}
In order to quantize the theory it is necessary to fix the gauge freedom, for instance with
a quadratic gauge-breaking term,
\begin{equation}
  S[\varphi]=S_{inv}[\varphi]+{1\over2}F^\alpha[\varphi]F_\alpha[\varphi].
\end{equation}
By performing a certain (non-local) gauge transformation on the path-integration variable $\varphi$
\cite{KKR,Fukuda:1976di},
it can be shown that an infinitesimal variation of the gauge fixing function, $\delta F^\beta$, induces
a change in the effective action as follows,
\begin{equation} \label{dGdX}
  \delta\Gamma[\bar\varphi]=-\Gamma_{,i}[\bar\varphi]\langle D^i_{\,\alpha}[\varphi]
  \mathcal {G}^\alpha_{\ \beta}[\varphi]\delta F^\beta[\varphi]\rangle[\bar\varphi]
  =:\Gamma_{,i}[\bar\varphi]\delta X^i[\bar\varphi],
\end{equation} 
where $\mathcal {G}^\alpha_{\ \beta}[\varphi]$ is the ghost propagator in a background field $\varphi$. 
This gauge dependence identity is a generalized version \cite{KKR,Fukuda:1976di} of the Nielsen 
identity \cite{Nielsen:1975fs,Aitchison:1984ns,Johnston:1987ib}.


\begin{thebibliography}{10}

\bibitem{Bailin:1984bm}
D. Bailin and A. Love, Phys. Rept. {\bf 107},  325  (1984).

\bibitem{Rajagopal:2000wf}
K. Rajagopal and F. Wilczek, The condensed matter physics of {QCD},
  hep-ph/0011333;
M.~G. Alford, Ann. Rev. Nucl. Part. Sci. {\bf 51},  131  (2001).

\bibitem{Son:1998uk}
D.~T. Son, Phys. Rev. {\bf D59},  094019  (1999).

\bibitem{Pisarski:1999tv}
R.~D. Pisarski and D.~H. Rischke, Phys. Rev. {\bf D61},  074017  (2000).

\bibitem{Schafer:1999jg}
T. Sch{\"a}fer and F. Wilczek, Phys. Rev. {\bf D60},  114033  (1999).

\bibitem{Hong:1999fh}
D.~K. Hong, V.~A. Miransky, I.~A. Shovkovy, and L.~C.~R. Wijewardhana, Phys.
  Rev. {\bf D61},  056001  (2000) [Erratum: ibid. {\bf D62}, 059903 (2000)].

\bibitem{Hsu:1999mp}
S.~D.~H. Hsu and M. Schwetz, Nucl. Phys. {\bf B572},  211  (2000).

\bibitem{Brown:1999aq}
W.~E. Brown, J.~T. Liu, and H.-c. Ren, Phys. Rev. {\bf D61},  114012  (2000).

\bibitem{Wang:2001aq}
Q. Wang and D.~H. Rischke, Phys. Rev. {\bf D65},  054005  (2002).

\bibitem{Rajagopal:2000rs}
K. Rajagopal and E. Shuster, Phys. Rev. {\bf D62},  085007  (2000).

\bibitem{Pisarski:2001af}
R.~D. Pisarski and D.~H. Rischke, Nucl. Phys. {\bf A702},  177  (2002).

\bibitem{Hong:2003ts}
D.~K. Hong {\it et~al.}, Higher order corrections to color superconducting
  gaps, hep-ph/0303181.

\bibitem{KKR}
R. Kobes, G. Kunstatter, and A. Rebhan, Phys. Rev. Lett. {\bf 64},  2992
  (1990);
Nucl. Phys. {\bf B355},  1  (1991);
A. Rebhan, Lect. Notes Phys. {\bf 583},  161  (2002).

\bibitem{Rischke:2000qz}
D.~H. Rischke, Phys. Rev. {\bf D62},  034007  (2000).

\bibitem{Pisarski:1999av}
R.~D. Pisarski and D.~H. Rischke, Phys. Rev. {\bf D60},  094013  (1999).

\bibitem{Manuel:2000nh}
C. Manuel, Phys. Rev. {\bf D62},  114008  (2000).

\bibitem{Casalbuoni:2003wh}
R. Casalbuoni and G. Nardulli, Inhomogeneous superconductivity in condensed
  matter and {QCD}, hep-ph/0305069.

\bibitem{Nielsen:1975fs}
N.~K. Nielsen, Nucl. Phys. {\bf B101},  173  (1975).

\bibitem{Aitchison:1984ns}
I.~J.~R. Aitchison and C.~M. Fraser, Ann. Phys. {\bf 156},  1  (1984).

\bibitem{LeB:TFT}
M. {Le Bellac}, {\em Thermal Field Theory} (Cambridge University Press,
  Cambridge, UK, 1996).


\bibitem{Baier:1992dy}
R. Baier, G. Kunstatter, and D. Schiff, Phys. Rev. {\bf D45},  4381  (1992).

\bibitem{Rebhan:1992ak}
A. Rebhan, Phys. Rev. {\bf D46},  4779  (1992).

\bibitem{Alford:1998zt}
M.~G. Alford, K. Rajagopal, and F. Wilczek, Phys. Lett. {\bf B422},  247
  (1998).

\bibitem{Amore:2001uf}
P. Amore, M.~C. Birse, J.~A. McGovern, and N.~R. Walet, Phys. Rev. {\bf D65},
  074005  (2002).

\bibitem{Alford:2002kj}
M. Alford and K. Rajagopal, JHEP {\bf 0206},  031  (2002).

\bibitem{Steiner:2002gx}
A.~W. Steiner, S. Reddy, and M. Prakash, Phys. Rev. {\bf D66},  094007  (2002).

\bibitem{Shovkovy:2003uu}
I. Shovkovy and M. Huang, Gapless two-flavor color superconductor,
  hep-ph/0302142.

\bibitem{Khlebnikov:1996vj}
S.~Y. Khlebnikov and M.~E. Shaposhnikov, Phys. Lett. {\bf B387},  817  (1996).

\bibitem{ItzZ:QFT}
C. Itzykson and J. Zuber, {\em Quantum Field Theory} (McGraw-Hill, New York,
  1985).


\bibitem{Fukuda:1976di}
R. Fukuda and T. Kugo, Phys. Rev. {\bf D13},  3469  (1976).

\bibitem{Johnston:1987ib}
D.~A. Johnston, Nucl. Phys. {\bf B283},  317  (1987).

\end{thebibliography}

\end{document}